\documentclass{article}

\usepackage[nonatbib, final]{neurips_ts4h_2022}

\usepackage[utf8]{inputenc} 
\usepackage[T1]{fontenc}    
\usepackage{hyperref}       
\usepackage{url}            
\usepackage{booktabs}       
\usepackage{amsfonts}       
\usepackage{nicefrac}       
\usepackage{microtype}      
\usepackage{xcolor}         
\usepackage{graphicx}
\usepackage{multirow}
\usepackage{bm}
\usepackage{subcaption}
\usepackage{xspace}
\usepackage{enumitem}
\usepackage{bold-extra}

\newcommand{\mname}{\texttt{NormIntSleep}\xspace}
\usepackage{subcaption}
\usepackage{enumitem}
\usepackage{float}
\usepackage{amsmath}

\title{Performance and utility trade-off in interpretable sleep staging}

\author{%
  Irfan Al-Hussaini\\
  Georgia Institute of Technology\\
  Atlanta, GA \\
  \texttt{alhussaini.irfan@gatech.edu} \\
   \And
  Cassie S. Mitchell\\
  Georgia Institute of Technology\\
  Atlanta, GA \\
  \texttt{cassie.mitchell@bme.gatech.edu} \\
}

\begin{document}

\maketitle

\begin{abstract}
Recent advances in deep learning have led to the development of models approaching the human level of accuracy. However, healthcare remains an area lacking in widespread adoption. The safety-critical nature of healthcare results in a natural reticence to put these black-box deep learning models into practice. This paper explores interpretable methods for a clinical decision support system called sleep staging, an essential step in diagnosing sleep disorders. Clinical sleep staging is an arduous process requiring manual annotation for each 30s of sleep using physiological signals such as electroencephalogram (EEG). Recent work has shown that sleep staging using simple models and an exhaustive set of features can perform nearly as well as deep learning approaches but only for some specific datasets. Moreover, the utility of those features from a clinical standpoint is ambiguous. On the other hand, the proposed framework, \mname demonstrates exceptional performance across different datasets by representing deep learning embeddings using normalized features. \mname performs 4.5\% better than the exhaustive feature-based approach and 1.5\% better than other representation learning approaches. An empirical comparison between the utility of the interpretations of these models highlights the improved alignment with clinical expectations when performance is traded-off slightly. \mname paired with a clinically meaningful set of features can best balance this trade-off by providing reliable, clinically relevant interpretation with robust performance. 



\end{abstract}

\section{Introduction}
There has been a steady accumulation of digital records of patient health data due to the increasingly widespread adoption of electronic health records \cite{jianxun2021electronic, adler2017hitech, jensen2012mining}. Breakthroughs in deep learning have leveraged this influx of clinical data to create increasingly complex and capable systems \cite{miotto2018deep, esteva2019guide, norgeot2019call}, which, coupled with optimized hardware, could lead to deployment in wearables on the edge \cite{athi_iedm, athi_jin, athi_ti1, athi_ti2}. However, the lack of interpretability and explainability of deep learning models prevents most of them from being used in practice because clinicians need to understand the reasoning behind each classification to avoid noise and bias \cite{zitnik, elshawi2019interpretability, carvalho2019machine, holzinger2019causability, wiens2019no}. Although challenging to design due to the manual effort required, linear models paired with a robust set of features can provide a degree of interpretability \cite{linear}. However, linear models paired with features too complex may result in models whose interpretation does not have clinical relevance. In this paper, using sleep staging as a case study, we take a deeper dive into the trade-off between the clinical relevance of explanations and performance and propose a model attempting to provide the ideal balance.

Around 70 million US adults are affected by sleep disorders  \cite{guglietta2015drug, holder2022common} such as insomnia, narcolepsy, or sleep apnea. The most crucial step for the diagnosis of sleep disorders is sleep staging \cite{RN50}. The gold standard for sleep staging remains the manual annotation of Polysomnogram (PSG) by clinicians  \cite{guillot2020dreem}. During this process, clinicians inspect the PSG signals from multiple channels and annotate each 30s segment with one of five sleep stages, i.e., wake, rapid eye movement (REM), and the non-REM stages N1, N2, and N3, by following guidelines stated in the American Academy of Sleep Medicine (AASM) Manual for the Scoring of Sleep and Associated Events \cite{berry2012aasm}. This manual annotation scheme is time-consuming and expensive because a clinician needs several hours to annotate a patient’s recordings from a single night \cite{ZHANG2022100371}.

There has been considerable research in automating sleep staging to overcome this problem. These approaches have primarily remained in the realm of deep learning that lack interpretability \cite{RN34}, for example, convolutional neural networks (CNN) \cite{RN49, RN57, RN9, yang2021single}, recurrent neural networks (RNN) \cite{RN14, phan2019seqsleepnet}, recurrent convolutional neural networks (RCNN) \cite{RN6, RN62}, graph convolutional networks (GCN) \cite{li2022attention, jia2020graphsleepnet}, and attention \cite{qu2020residual, phan2022sleeptransformer, li2022attention}. On the other hand, the AASM sleep scoring manual guidelines \cite{berry2012aasm} are interpretable for clinicians but need more explicit definitions to design a robust computational model \cite{al2019sleeper}.

A recent study proposed a feature-based linear model \cite{linear} that performed as well as deep neural networks. However, the features used in this study were not designed considering clinical guidelines. To generate clinically relevant explanations, we propose \mname, a representation learning framework that projects deep neural network embeddings into an interpretable normalized feature space. Thus, by choosing an appropriate feature space, \mname can unite clinically relevant explanations with the high accuracy of a deep learning model.

\section{Data}
\label{sec:data}
Two publicly available datasets are used for evaluation. These datasets are summarized in Table \ref{tab:dataset}. In PhysioNet-EDFX \cite{physionet1, physionet2, goldberger2000physiobank}, sleep stages N3 and N4 annotated using the R\&K schema \cite{rk_1, rk_2} were combined into a single N3 class to align with the AASM standards \cite{berry2012aasm, moser2009sleep, danker2009interrater} used in the ISRUC dataset \cite{isruc}. After this alignment procedure, both datasets provide a sleep stage annotation from wake (W), rapid eye movement (REM), and the non-REM stages (N1, N2, N3) to each 30-second epoch. Interpretable sleep staging aims to predict these sleep stages using multi-channel physiological signals while providing clinically meaningful interpretations for each classified sleep stage.

\section{Method}
\label{sec:method}

\begin{table}[tb]
  \begin{minipage}{\linewidth}
    \centering
\caption[for LOF]{Datasets}
\resizebox{\linewidth}{!}{%
\begin{tabular}{lcccc}
\hline
                        & \begin{tabular}[c]{@{}c@{}}Number of \\ Subjects\end{tabular} & \begin{tabular}[c]{@{}c@{}}Sampling \\ Frequency (Hz)\end{tabular} & \begin{tabular}[c]{@{}c@{}}Channel \\ Names\end{tabular} & \begin{tabular}[c]{@{}c@{}}Annotation \\ Schema\end{tabular} \\ \hline
ISRUC \cite{RN23}          & 100                                                                    & 200                                                                         & \begin{tabular}[c]{@{}c@{}}F3-A2, C3-A2, F4-A1, C4-A1, O1-A2,\\ O2-A1, ROC-A1, LOC-A2, Chin-EMG\end{tabular}                                                                      & AASM \cite{berry2012aasm} \\ 
PhysioNet-EDFX \cite{physionet1, physionet2} & 197                                                                    & 100                                                                         & \begin{tabular}[c]{@{}c@{}}EEG Fpz-Cz, EEG Pz-Oz,\\ EOG horizontal, EMG submental\end{tabular}                                                                      & R\&K \cite{rk_1, rk_2}                       \\ \hline
\end{tabular}}
\label{tab:dataset}
\end{minipage}
\end{table}

\mname uses the PSG recordings, $\bm{X}$, to generate an interpretable representation for deep neural network embeddings using the following steps:
\begin{enumerate}
\item A CNN-LSTM network \cite{serf} is trained end-to-end on sleep staging using the multi-channel EEG, EOG, and EMG signals as input. The CNN is composed of 3 convolutional layers where each layer is followed by batch normalization, ReLU activation, and max pooling. The kernel sizes of the three layers are 201, 11, and 11, and the output channels are 256, 128, and 64. The CNN output is used as input for a layer of bi-directional Long Short-Term Memory (LSTM) cells with 256 hidden states. The resulting 512 hidden states represent the \textit{embedding space}, $\mathcal{E}$. During model training using cross-entropy loss, the LSTM output, $\bm{E(X)}$ is connected to a fully-connected layer with 5 outputs for the 5 sleep stages.
\item Features $\bm{F(X)}$, defined in the \textit{feature space} $\mathcal{F}$, are extracted from the dataset. There are two possible sets of features:
    \begin{itemize}
        \item \textbf{FeatLong}: an exhaustive list of features inspired by the recent work of Van Der Donckt et al. \cite{linear}. The features are not designed considering clinical guidelines in the AASM Manual for the Scoring of Sleep and Associated Events \cite{berry2012aasm}. It results in 2488 features for the ISRUC dataset and 1048 features for the Physionet dataset. 10\% of the most significant features are retained for the next steps using ANOVA, resulting in 249 features for the ISRUC dataset and 105 for the Physionet dataset.
        \item \textbf{FeatShort}: a smaller set of clinically interpretable features inspired by the recent work of Al-Hussaini et al. \cite{serf}. The features are designed according to the AASM manual \cite{berry2012aasm}. 87 features are extracted for the ISRUC dataset and 38 for the Physionet dataset. 90\% of the most significant features are retained for the next steps using ANOVA, resulting in 78 features for the ISRUC dataset and 34 for the Physionet dataset.
    \end{itemize}
\item A linear transformation, $\bm{T}$, is learned from the embedding space to the feature space, $\mathcal{E} \xrightarrow[]{\bm{T}} \mathcal{F}$, using linear least squares regression with $L_2$ regularization. $\bm{R} = \bm{E(X)} \cdot \bm{T}$ defines the interpretable representations of embeddings after projecting the embedding, $\bm{E(X)}$,  to the interpretable feature space, $\mathcal{F}$. These representations of the embeddings, $\bm{R}$, are normalized before being used as input to classifiers.
    $$\bm{R'}=\frac{\bm{R}-\mu_R}{\sigma_R}$$
    where $\mu_R = \frac{\sum R}{N}$ and $\sigma_R = \sqrt{\frac{1}{N} \sum_{i=1}^N (R_i - \mu_R)^2}$.
    
These normalized interpretable representations of the embeddings, $\bm{R'}$, are used as inputs to simple classifiers such as logistic regression and decision tree.
\end{enumerate}

\section{Experiments}
\label{sec:experiments}
\subsection{Experimental setup}
\label{sec:setup}
The CNN-LSTM network was trained using PyTorch 1.0 \cite{NEURIPS2019_bdbca288} and a batch size of 1000 samples from 1 PSG. The training was continued for 20 epochs with a learning rate of $10^{-4}$ using ADAM \cite{RN25} as the optimization method. The simple models were trained using scikit-learn \cite{RN44, cuml}, XGBoost \cite{xgboost}, and CatBoost \cite{catboost1, catboost2}. The features were extracted using scikit-learn \cite{RN44, cuml}, yasa \cite{yasa}, and tsflex \cite{vanderdonckt2021tsflex}. 
The data is randomly split by subjects into a training and a test set in a 9:1 ratio using the same seed for all experiments. The training set is used for each dataset to fix model parameters, and the test set is used to obtain performance metrics. The same model hyperparameters and feature extraction schema are used to prevent overfitting and ensure consistent performance across different datasets.

\subsection{Baselines}
\begin{itemize}[leftmargin=*]
	\item CNN-LSTM: the model used to generate the embeddings 
	\item FeatLong and FeatShort: defined in Section \ref{sec:method}. They are used as inputs to simple classifiers - CatBoost \cite{catboost1,catboost2}, XGBoost \cite{xgboost}, Logistic Regression, and Gradient Boosted Trees
	\item SLEEPER \cite{al2019sleeper}: a prototype based interpretable sleep staging algorithm
	\item SERF \cite{serf}: an interpretable sleep staging algorithm based on embeddings, rules, and features
	\item U-Time \cite{RN57}: state-of-the-art deep learning model for sleep staging
\end{itemize}
\subsection{Results}
\begin{table}[htb]
  \begin{minipage}{\linewidth}
    \centering
\caption[LOR]{Model Evaluation}
\label{tab:eval}
\resizebox{\linewidth}{!}{%
\begin{tabular}{l@{\qquad}cc@{\qquad}cc@{\qquad}cc@{\qquad}c}
  \toprule
  \multirow{2}{*}{\raisebox{-\heavyrulewidth}{Model}} & \multicolumn{2}{c}{Accuracy}  & \multicolumn{2}{c}{F1 Score (Macro)} & \multicolumn{2}{c}{Cohen's $\kappa$} & \multirow{2}{*}{\raisebox{-\lightrulewidth}{\begin{tabular}[c]{@{}c@{}}Average\\Performance\end{tabular}}}\\
  \cmidrule{2-7}
                                           & Physionet       & ISRUC      & Physionet          & ISRUC         & Physionet     & ISRUC  &   \\
                                           \midrule
\mname-FeatLong-XGBoost         & 0.847           & 0.810      & 0.798              & 0.785         & 0.789         & 0.754     & 0.797                           \\
FeatLong-XGBoost \cite{linear}                & 0.836           & 0.750      & 0.777              & 0.712         & 0.773         & 0.675     & 0.754                           \\
\mname-FeatLong-CatBoost   & 0.846           & 0.815      & 0.793              & 0.788         & 0.787         & 0.760     & 0.798                           \\
FeatLong-CatBoost \cite{linear}         & 0.834           & 0.743      & 0.766              & 0.701         & 0.768         & 0.666     & 0.746                           \\
\underline{\mname-FeatLong-Logistic Regression}       & 0.855           & 0.807      & 0.801              & 0.783         & 0.800         & 0.748     & \underline{0.799}                           \\
FeatLong-Logistic Regression \cite{linear}               & 0.809           & 0.760      & 0.736              & 0.717         & 0.733         & 0.689     & 0.741                           \\
\mname-FeatShort-XGBoost        & 0.832           & 0.809      & 0.772              & 0.781         & 0.768         & 0.753     & 0.786                           \\
FeatShort-XGBoost               & 0.825           & 0.790      & 0.767              & 0.747         & 0.759         & 0.726     & 0.769                           \\
\mname-FeatShort-CatBoost  & 0.831           & 0.809      & 0.770              & 0.777         & 0.766         & 0.752     & 0.784                           \\
FeatShort-CatBoost         & 0.818           & 0.788      & 0.750              & 0.741         & 0.747         & 0.724     & 0.761                           \\
\underline{\mname-FeatShort-Logistic Regression}      & 0.855           & 0.797      & 0.799              & 0.774         & 0.800         & 0.735     & \underline{0.793}                           \\
SERF-XGBoost  \cite{serf}                  & 0.823           & 0.819      & 0.753              & 0.789         & 0.753         & 0.766     & 0.784                           \\
SERF-Logistic Regression  \cite{serf}                  & 0.829           & 0.795      & 0.759              & 0.773         & 0.762         & 0.733     & 0.775                           \\
SLEEPER-Gradient Boosted Trees  \cite{al2019sleeper}               & 0.807           & 0.797      & 0.721              & 0.756         & 0.729         & 0.736     & 0.758                           \\
\textbf{U-Time}  \cite{RN57}                   & 0.862           & 0.840      & 0.811              & 0.816         & 0.810         & 0.793     & \textbf{0.822}                           \\
CNN-LSTM  \cite{serf}                 & 0.864           & 0.831      & 0.815              & 0.819         & 0.813         & 0.783     & 0.821       \\
\bottomrule
\end{tabular}}
\end{minipage}
\end{table}

Accuracy, Macro F1-Score, and Cohen's $\kappa$ are used to evaluate the models. The results in Table \ref{tab:eval} were produced in the same experimental setup expanded upon in Section \ref{sec:setup}. Since the performance of models varies a lot between the two datasets, an aggregated metric, \textit{Average Performance}, is calculated based on the average value of each model across the two datasets and three metrics. The best interpretable methods using FeatShort and FeatLong are underlined, and the best overall approach is highlighted in bold. The results show that \mname surpasses all other interpretable methods, even when it uses the smaller set of interpretable features, FeatShort. The benefits of using this over the exhaustive set of features, FeatLong, are discussed in Section \ref{sec:interpretability}.

\subsection{Interpretation and Clinical Relevance}
\label{sec:interpretability}

\begin{figure*}[htb]
    \centering
    \begin{subfigure}[b]{\textwidth}
    {
        \centering
        \includegraphics[width=\textwidth, keepaspectratio]{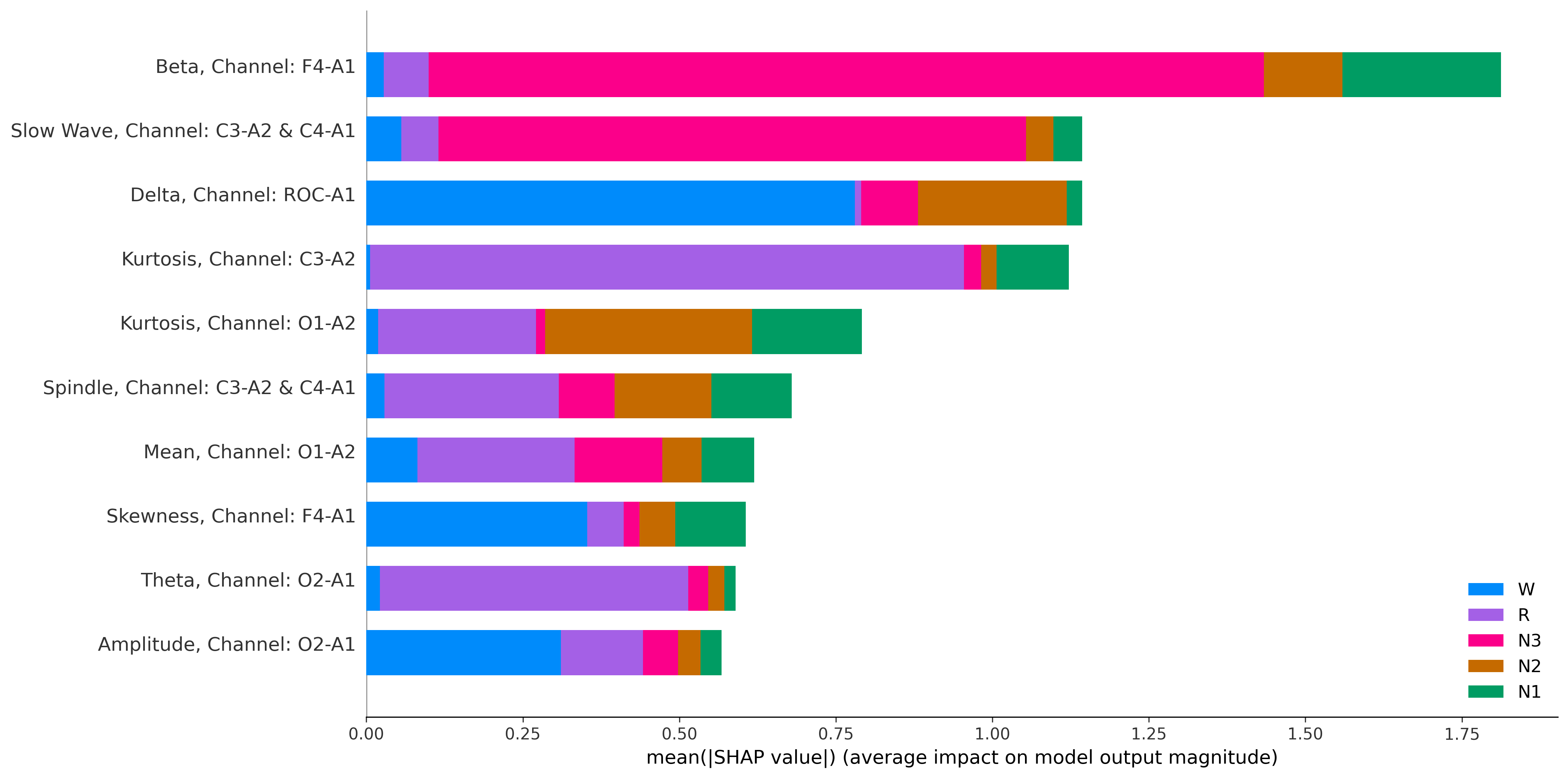}
        \caption{\mname-XGBoost-FeatShort}
        \label{fig:shap_sim_xgboost_old}
    }
    \end{subfigure}
    \begin{subfigure}[b]{\textwidth}
    {
        \centering
        \includegraphics[width=\textwidth]{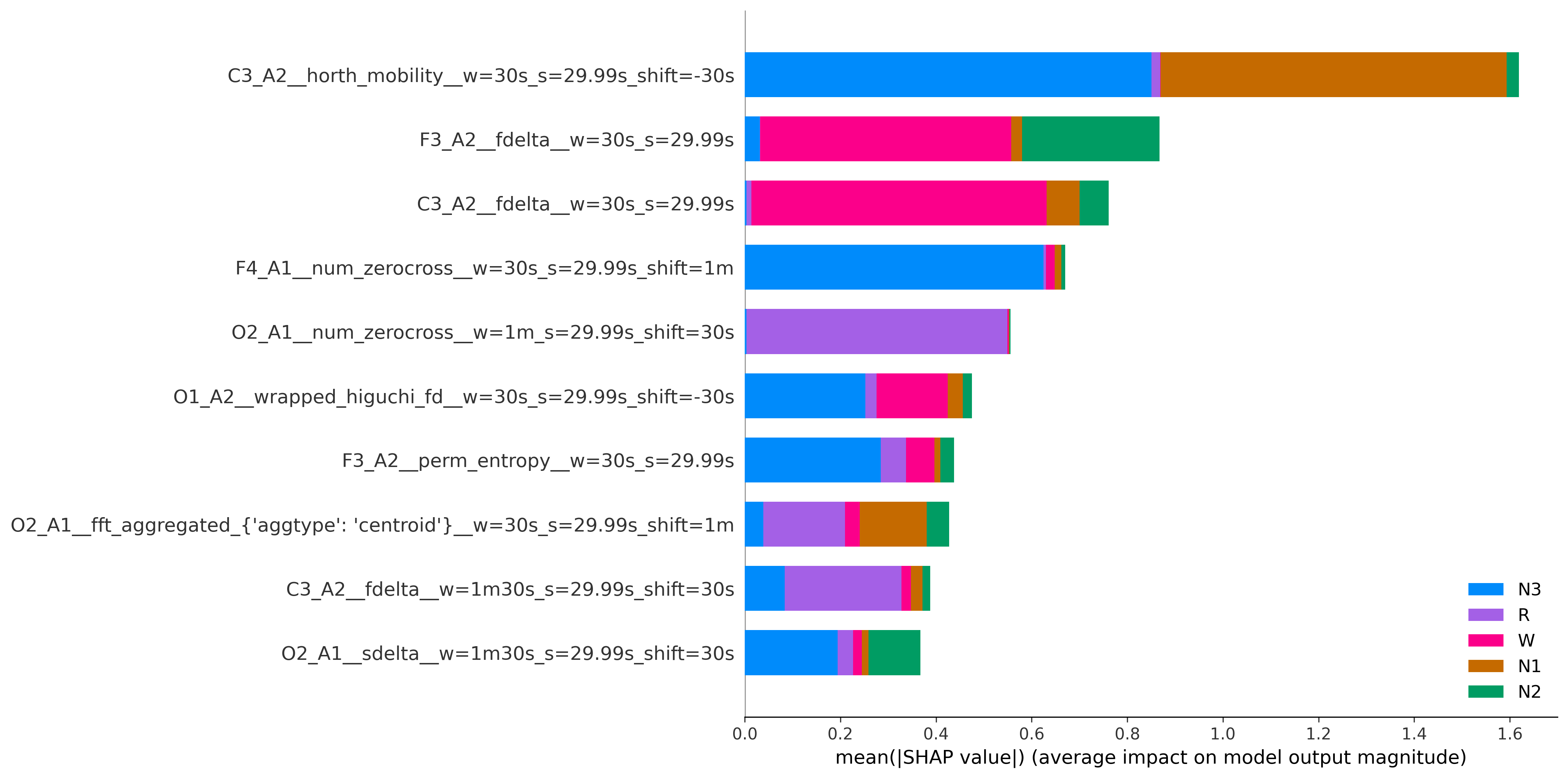}
        \caption{\mname-XGBoost-FeatLong}
        \label{fig:shap_sim_xgboost_imec}
    }
    \end{subfigure}
\caption{The most important embedding representations in \mname and the influence on each sleep stage classification according to SHAP values (ISRUC Dataset)}
\label{fig:shap_sim_main}
\end{figure*}

Figure \ref{fig:shap_sim_main} highlights the top 10 important dimensions of the interpretable representation of the \mname embeddings, $\bm{R'}$, according to SHAP values and their influence on the classification of each of the 5 sleep stages. N1 is a challenging sleep stage to classify, with around 50\% agreement among human annotators, so its classification is not discussed in detail. The differences between Figure \ref{fig:shap_sim_xgboost_old} and Figure \ref{fig:shap_sim_xgboost_imec} show the importance of having clinically relevant representations. A deeper analysis is performed on the 3 most critical interpretable representations of embeddings in \mname-XGBoost-FeatShort (Figure \ref{fig:shap_sim_xgboost_old}): 
\begin{enumerate}
    \item  Beta waves \cite{serf} have frequencies between 8 Hz and 20 Hz. This frequency range is present in Wake \cite{beta}, REM \cite{beta}, and N2 (through the overlapping sigma band in Spindles \cite{spindles_sigma}). Consequently, it is the perfect attribute to differentiate N3 from the other stages and aligns with the high SHAP value assigned to N3.
    \item  Slow waves are essential for restorative sleep \cite{dijk2009regulation}. As a result, slow waves play a critical role in maintaining sleep and daytime function in patients with insomnia or nonrestorative sleep \cite{dijk2009regulation}. During sleep staging, these slow waves are used by clinicians to annotate N3 \cite{beta, berry2012aasm}. Thus, clinical evidence supports the importance of slow waves in N3 classification by \mname-XGBoost-FeatShort.
    \item EOG has been used to detect a person's wakefulness \cite{ganesan2020binary} and movement of the eyes \cite{lin2019eog, barea2012eog}. Torsvall et al. \cite{torsvall1988extreme} show that Delta activity in EOG is highest in the wake stage before sleep induction. Thus, the high SHAP value attributed to Delta waves in ROC-A1 (an EOG channel) in classifying wake using \mname-XGBoost-FeatShort aligns with clinical expectations. 
\end{enumerate}

The alignments with clinical guidelines highlight the utility of the explanations provided by \mname-XGBoost-FeatShort. On the other hand, 6 of the top 10 representations in \mname-XGBoost-FeatLong (Figure \ref{fig:shap_sim_xgboost_imec}) are not defined using clinical guidelines such as the AASM manual for the scoring of sleep \cite{berry2012aasm}, and the rest are derived from the same frequency band, delta. Thus, the clinical relevance of \mname-XGBoost-FeatLong's interpretations is ineffective. The sleep staging interpretations such as that provided by \mname-XGBoost-FeatShort can give physicians confidence in the automated labels and facilitate validation, thus paving the way for faster adoption.

The feature space of the two datasets using FeatLong and FeatShort are shown in Appendix \ref{app:feat_viz}. It highlights the efficacy of FeatLong in segregating the Physionet dataset into the 5 classes through the well-partitioned clusters for each sleep stage. The importance of features using XGBoost paired with FeatShort and FeatLong are shown in Appendix \ref{app:feat_imp}.

\section{Conclusion}
Interpretability is crucial for the adoption of clinical decision support systems. Complex features paired with a simple model can provide interpretation. However, those explanations can only be helpful if the features are clinically meaningful. \mname offers a generalizable framework to leverage the most clinically significant features for classification with accuracy higher than complex feature-based models and similar to black-box deep learning. As a result, the interpretations are clinically relevant. Thus \mname takes forward strides towards adoption.
\section*{Acknowledgment}
This research was funded by the National Science Foundation CAREER grant 1944247 to C.M, the National Institute of Health grant U19-AG056169 sub-award to C.M., and the McCamish Parkinson’s Disease Innovation Program at Georgia Institute of Technology and Emory University to C.M.

\bibliographystyle{unsrt}
\bibliography{refs}
\appendix
\clearpage
\section{Appendix A: Feature Space Visualization} 
\label{app:feat_viz}
\begin{figure*}[htb]
    \centering
    \begin{subfigure}[b]{0.5\textwidth}
    {
        \centering
        \includegraphics[width=\textwidth, keepaspectratio]{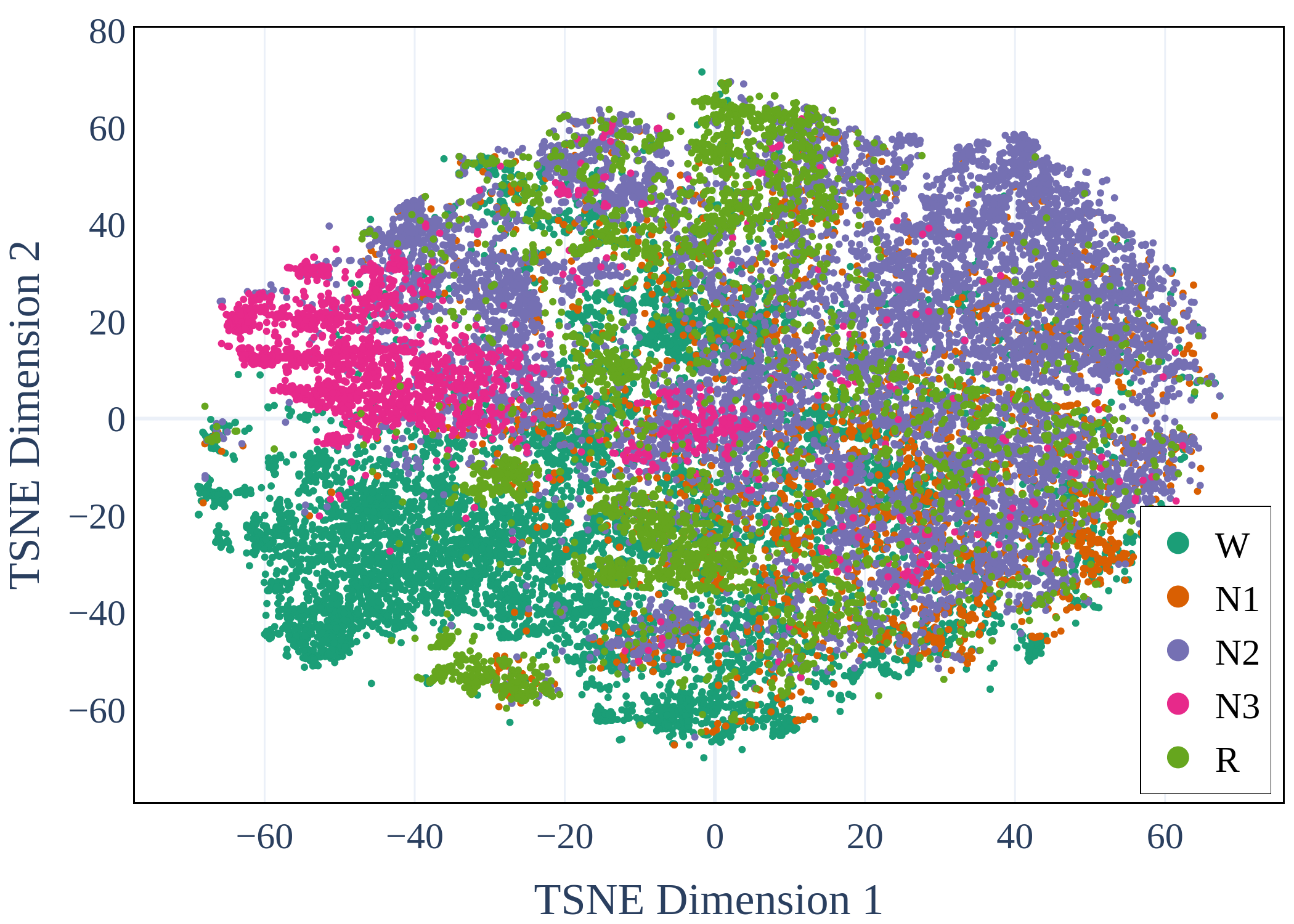}
        \caption{TSNE using FeatShort}
        \label{fig:tsne_less_physionet}
    }
    \end{subfigure}%
    \begin{subfigure}[b]{0.5\textwidth}
    {
        \centering
        \includegraphics[width=\textwidth]{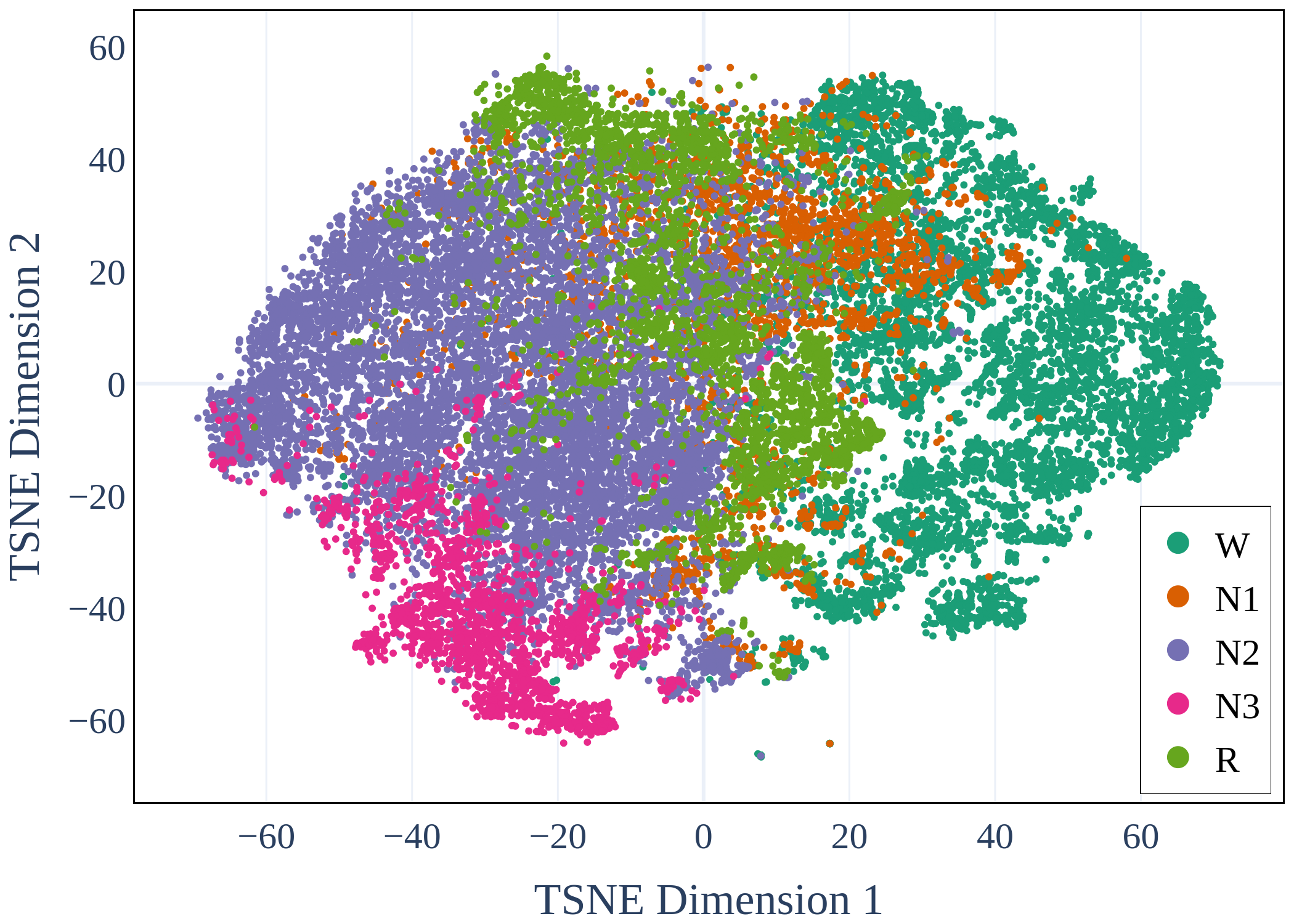}
        \caption{TSNE using FeatLong}
        \label{fig:tsne_more_physionet}
    }
    \end{subfigure}
    \vspace{0.5em}
    \begin{subfigure}[b]{0.5\textwidth}
    {
        \centering
        \includegraphics[width=\textwidth, keepaspectratio]{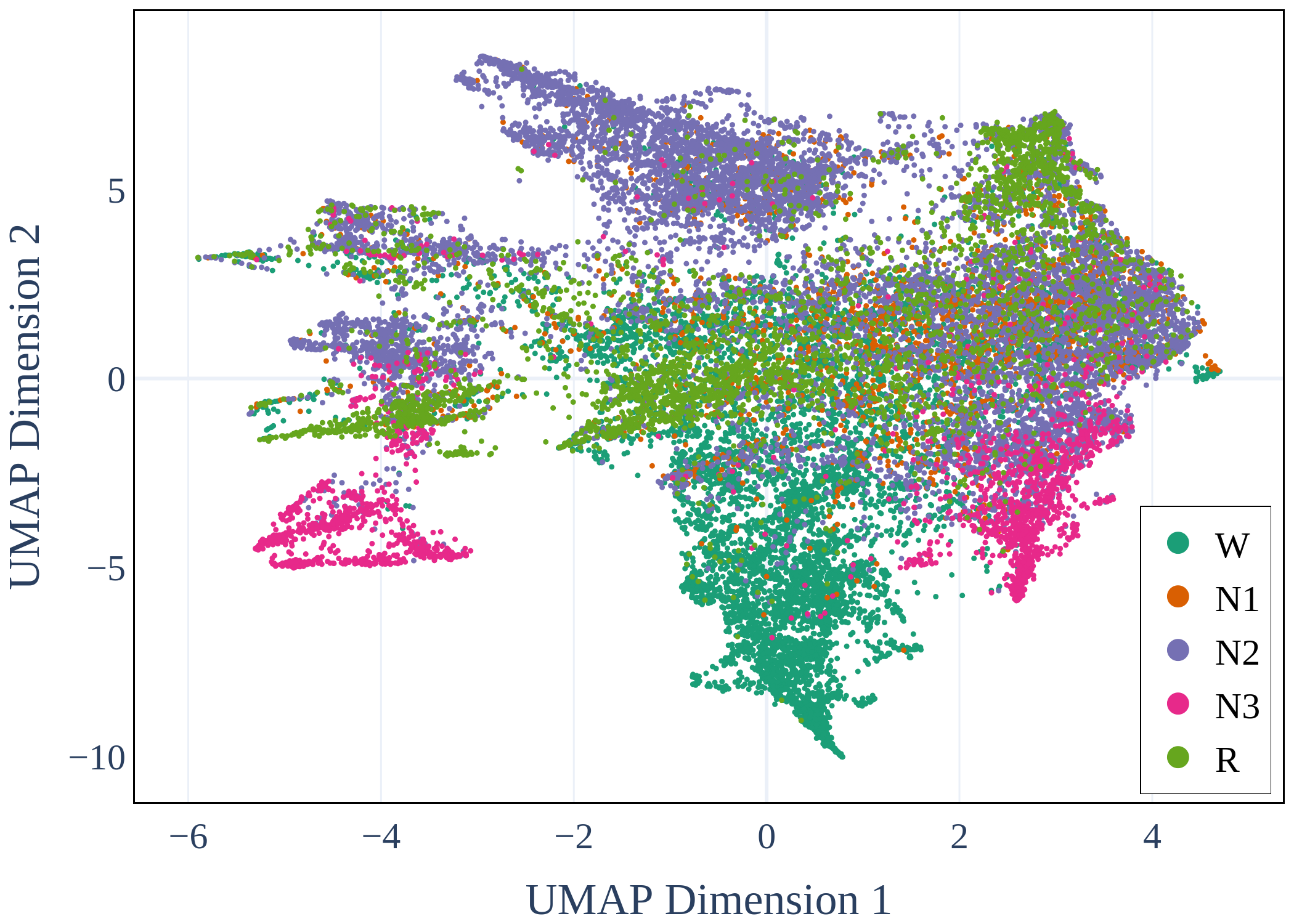}
        \caption{UMAP using FeatShort}
        \label{fig:umap_less_physionet}
    }
    \end{subfigure}%
    \begin{subfigure}[b]{0.5\textwidth}
    {
        \centering
        \includegraphics[width=\textwidth]{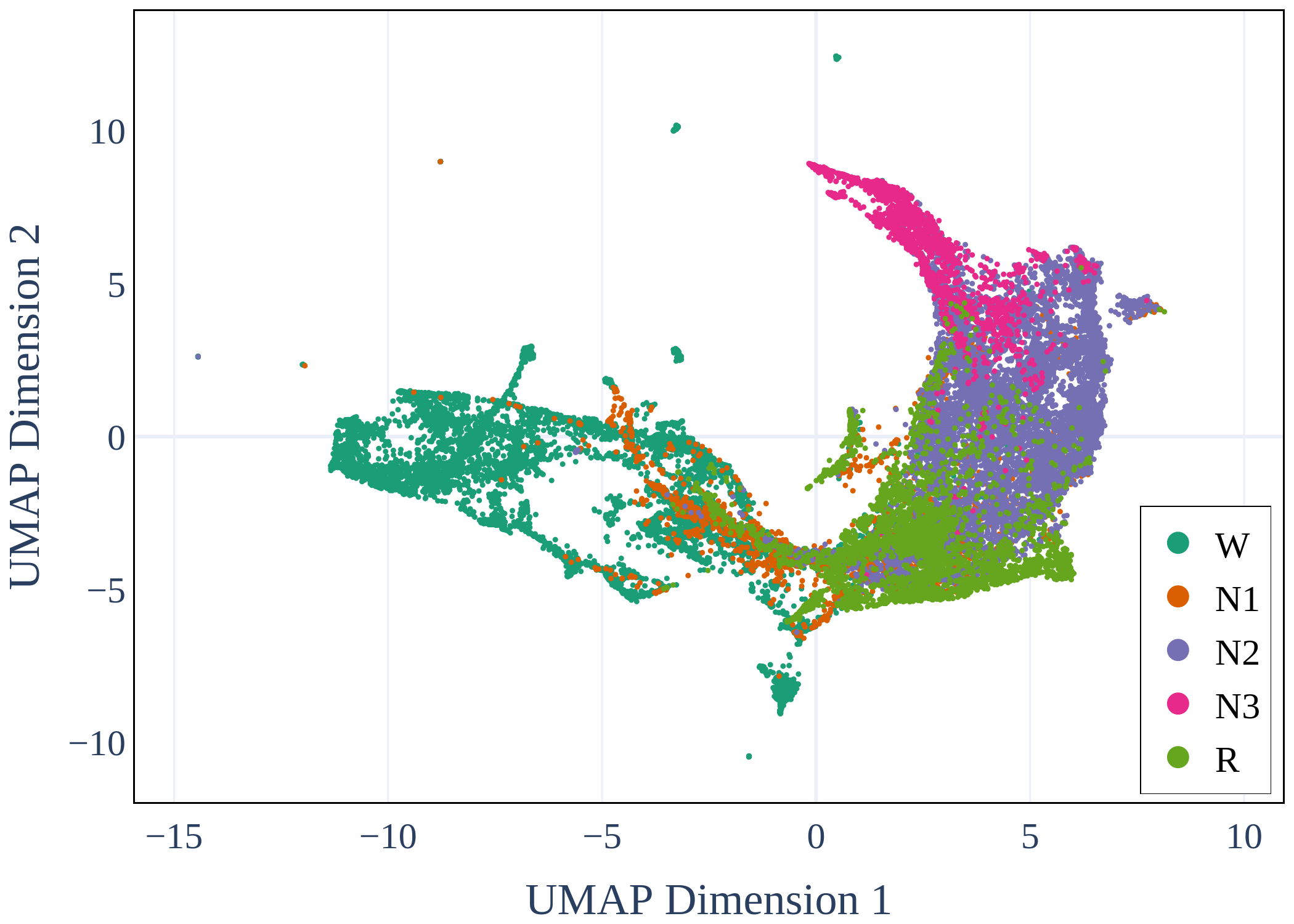}
        \caption{UMAP using FeatLong}
        \label{fig:umap_more_physionet}
    }
    \end{subfigure}
\caption{Dimensionality reduction on the Physionet dataset: shows distinct clusters for classes using FeatLong}
\label{fig:dim_red_physionet}
\end{figure*}

\clearpage
\begin{figure*}[htb]
    \centering
    \begin{subfigure}[b]{0.5\textwidth}
    {
        \centering
        \includegraphics[width=\textwidth, keepaspectratio]{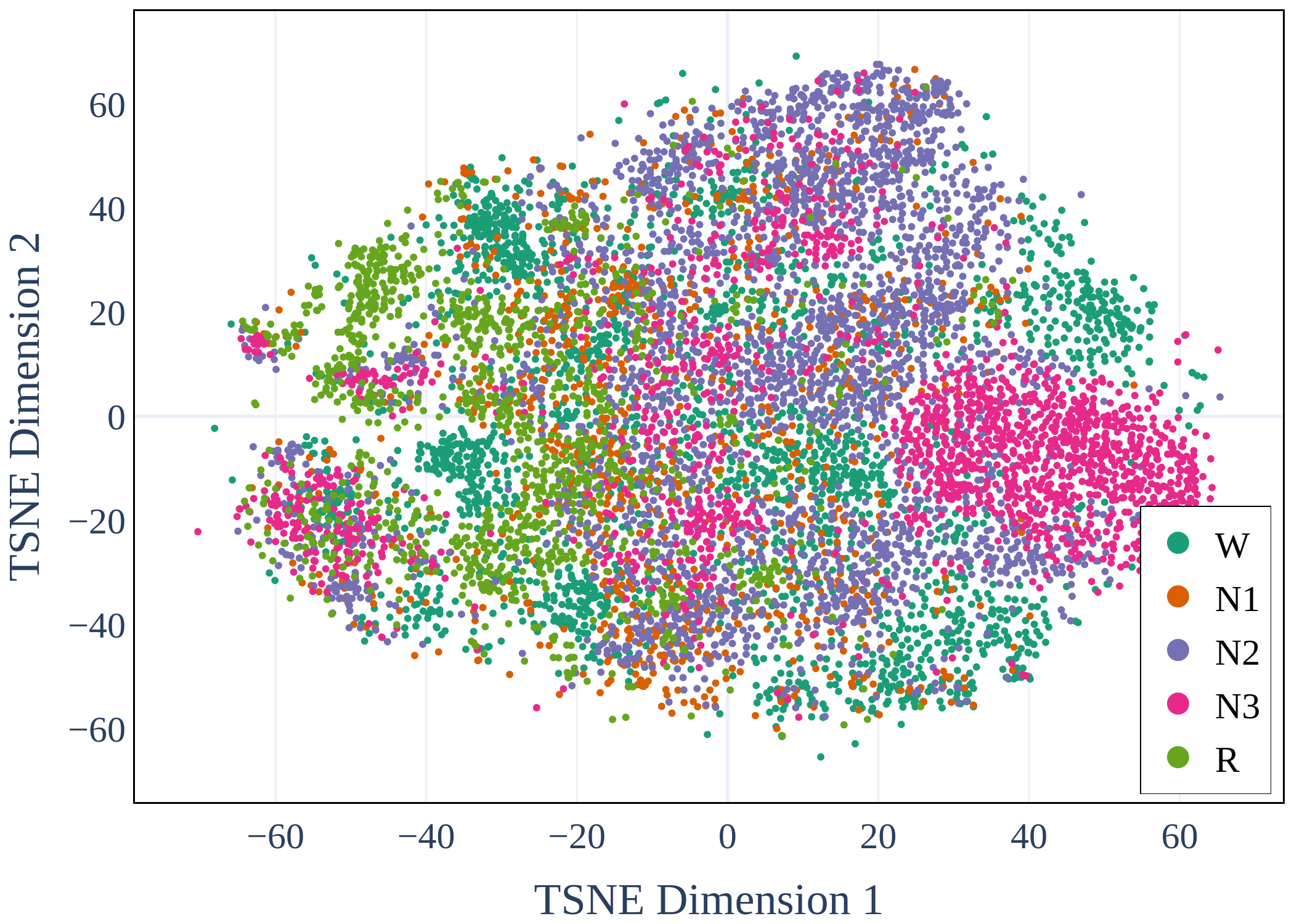}
        \caption{TSNE using FeatShort}
        \label{fig:tsne_less}
    }
    \end{subfigure}%
    \begin{subfigure}[b]{0.5\textwidth}
    {
        \centering
        \includegraphics[width=\textwidth]{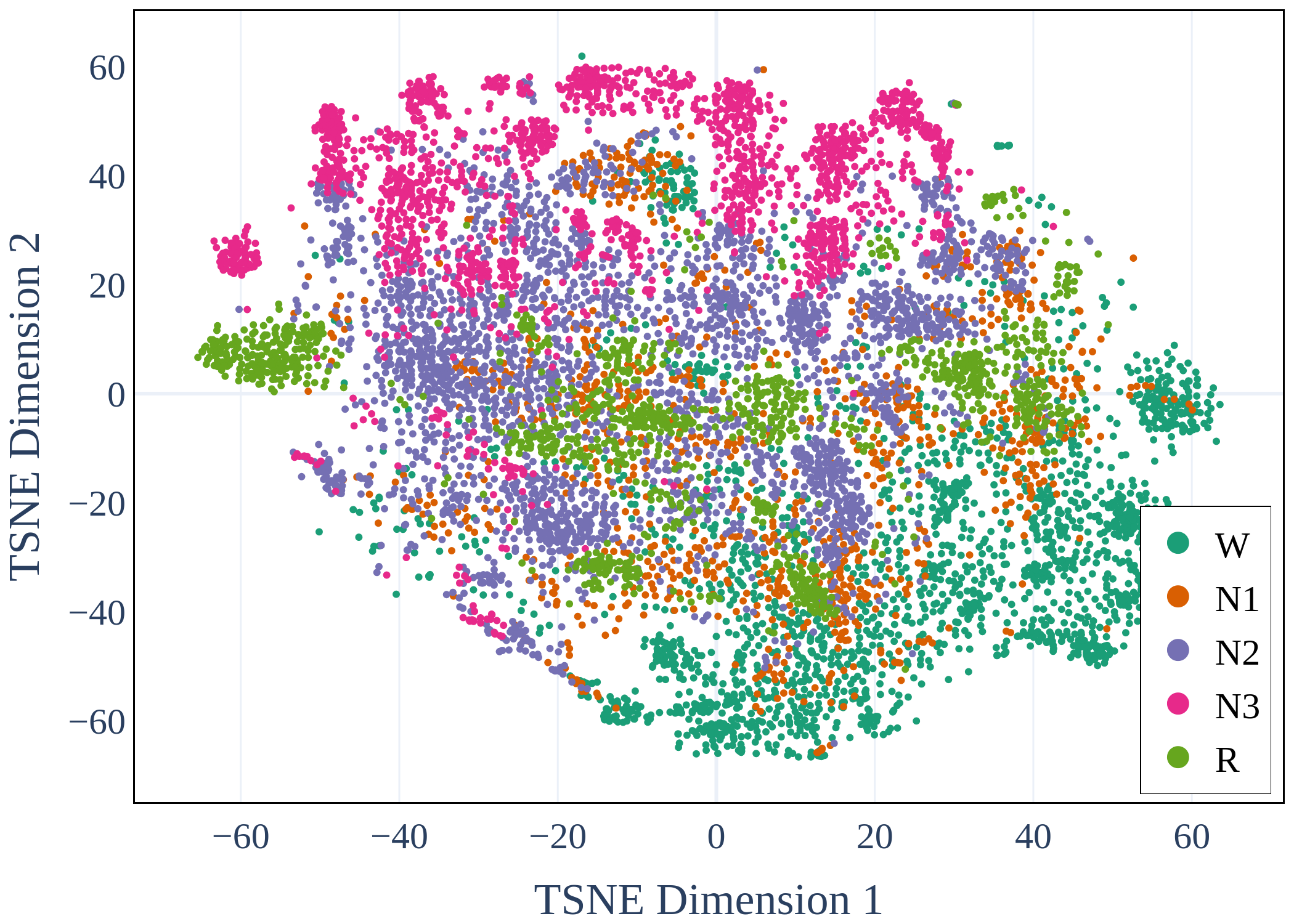}
        \caption{TSNE using FeatLong}
        \label{fig:tsne_more}
    }
    \end{subfigure}
    \vspace{0.5em}
    \begin{subfigure}[b]{0.5\textwidth}
    {
        \centering
        \includegraphics[width=\textwidth, keepaspectratio]{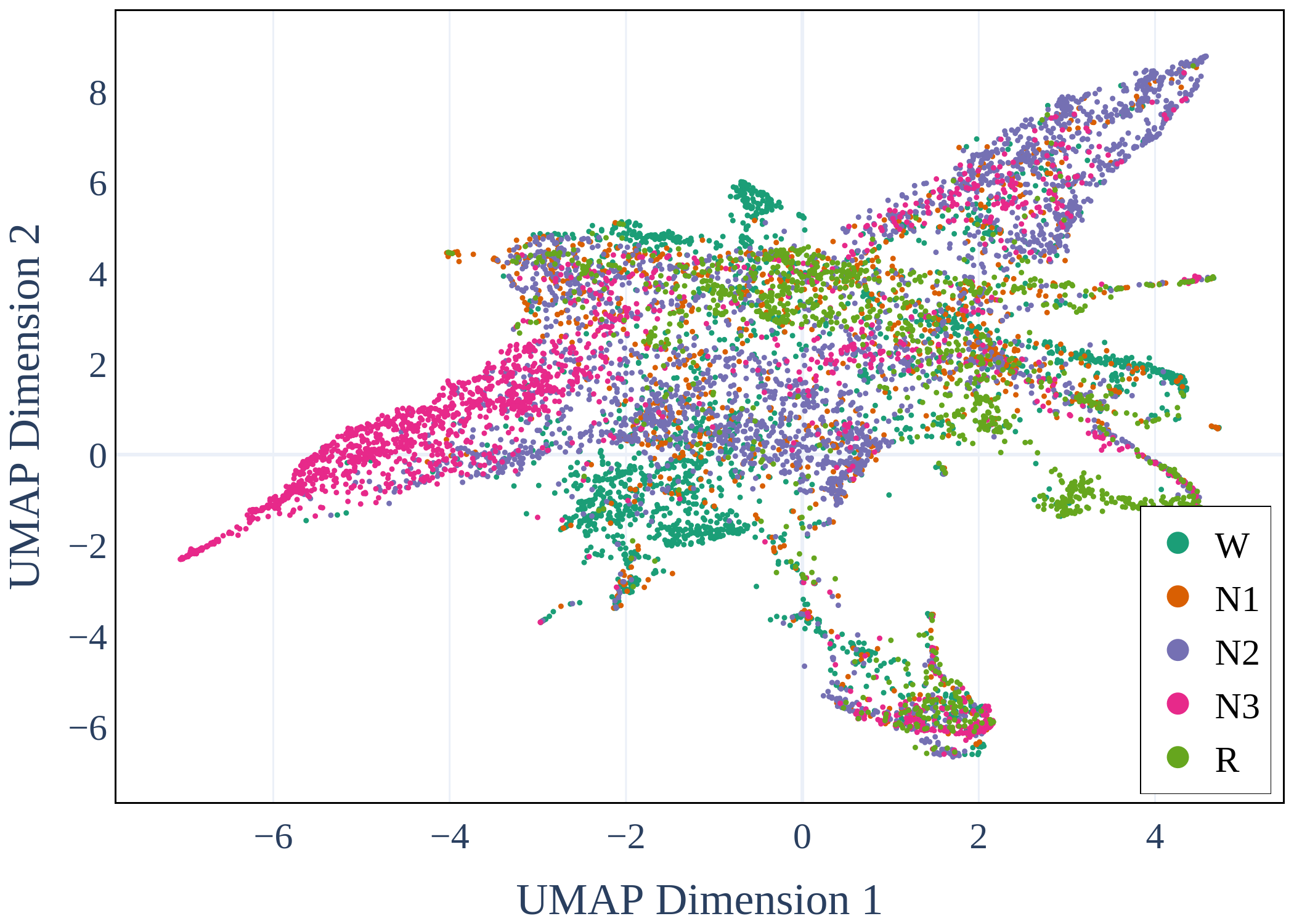}
        \caption{UMAP using FeatShort}
        \label{fig:umap_less}
    }
    \end{subfigure}%
    \begin{subfigure}[b]{0.5\textwidth}
    {
        \centering
        \includegraphics[width=\textwidth]{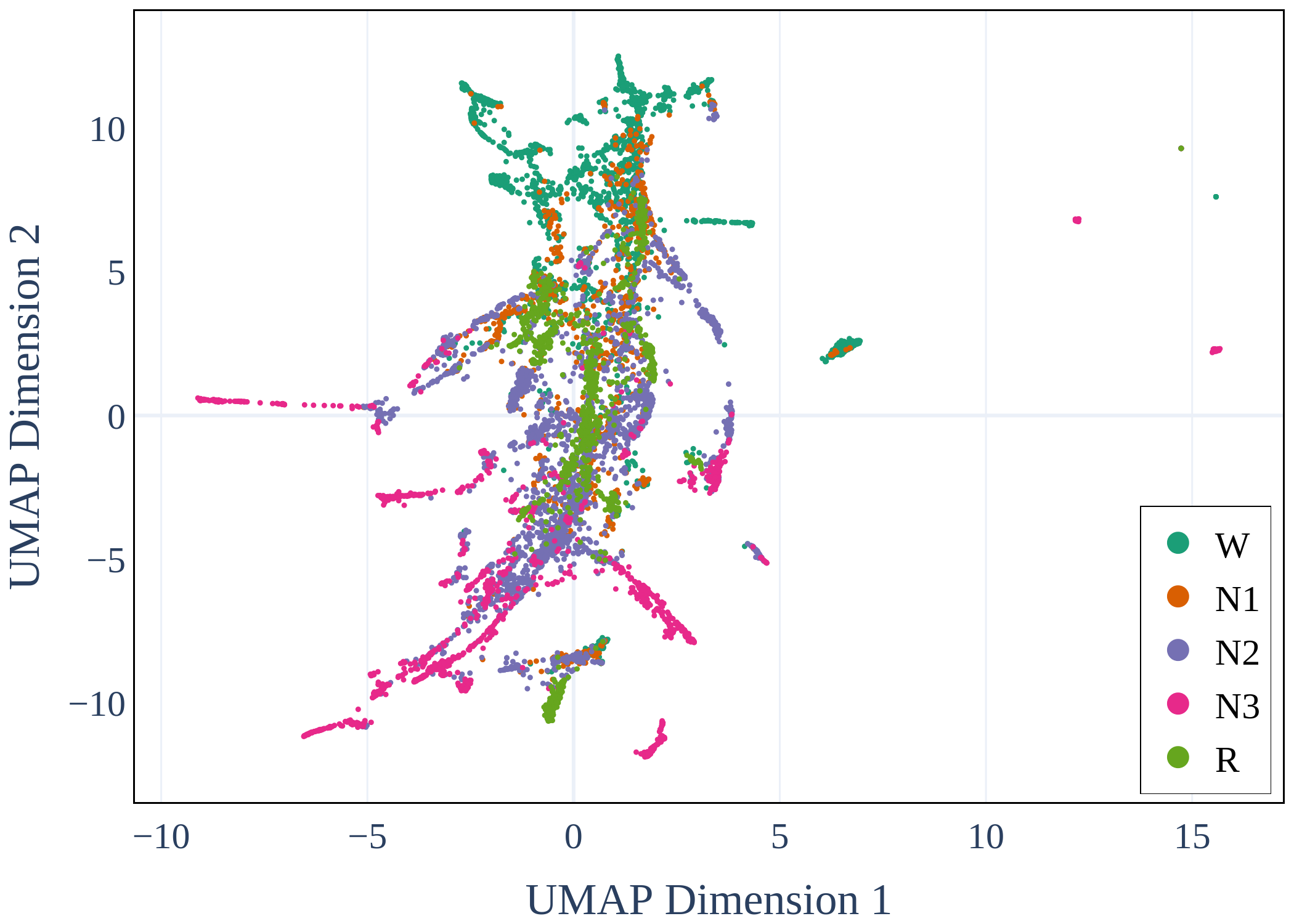}
        \caption{UMAP using FeatLong}
        \label{fig:umap_more}
    }
    \end{subfigure}
\caption{Dimensionality reduction on the ISRUC dataset}
\label{fig:dim_red}
\end{figure*}
\clearpage
\section{Appendix B. SHAP Feature Importance} 
\label{app:feat_imp}

\begin{figure*}[htb]
    \centering
    \begin{subfigure}[b]{0.9\textwidth}
    {
        \centering
        \includegraphics[width=\textwidth, keepaspectratio]{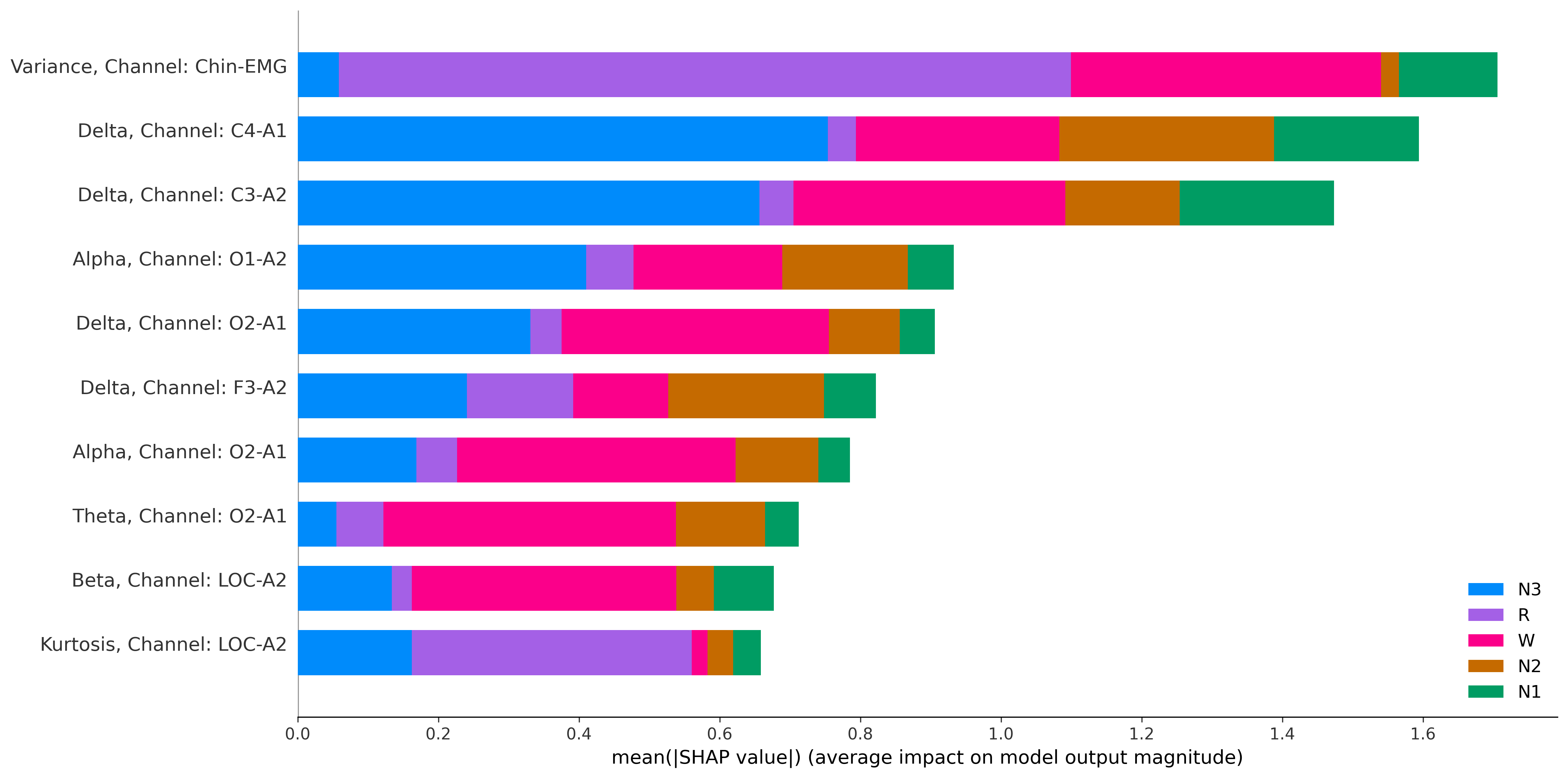}
        \caption{XGBoost-FeatShort}
        \label{fig:shap_feat_xgboost_old}
    }
    \end{subfigure}
    \begin{subfigure}[b]{0.9\textwidth}
    {
        \centering
        \includegraphics[width=\textwidth]{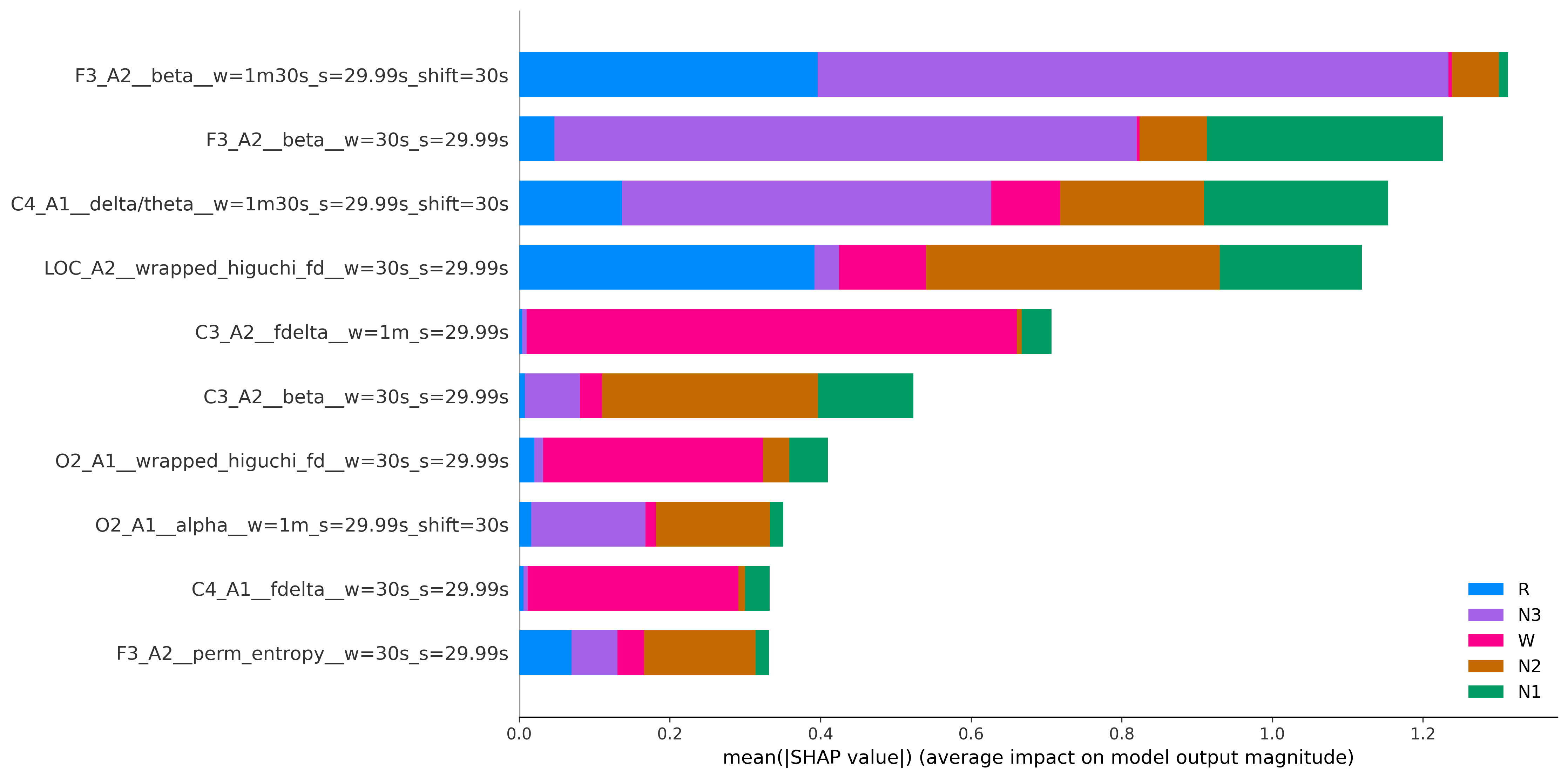}
        \caption{XGBoost-FeatLong}
        \label{fig:shap_feat_xgboost_imec}
    }
    \end{subfigure}
\caption{SHAP feature importance using feature-based models on ISRUC Dataset}
\label{fig:shap_feat}
\end{figure*}



\end{document}